\documentstyle[11pt,epsfig,aaspp4]{article}
\tighten 
\begin{document}

\baselineskip 5.0ex

\newcommand{\lya}{Lyman~$\alpha$}
\newcommand{\lyb}{Lyman~$\beta$}
\newcommand{\za}{$z_{\rm abs}$}
\newcommand{\ze}{$z_{\rm em}$}
\newcommand{\cmtwo}{cm$^{-2}$}
\newcommand{\nhi}{$N$(H$^0$)}
\newcommand{\nzn}{$N$(Zn$^+$)}
\newcommand{\ncr}{$N$(Cr$^+$)}
\newcommand{\degpoint}{\mbox{$^\circ\mskip-7.0mu.\,$}}
\newcommand{\halpha}{\mbox{H$\alpha$}}
\newcommand{\hbeta}{\mbox{H$\beta$}}
\newcommand{\hgamma}{\mbox{H$\gamma$}}
\newcommand{\kms}{\,km~s$^{-1}$}      
\newcommand{\minpoint}{\mbox{$'\mskip-4.7mu.\mskip0.8mu$}}
\newcommand{\mv}{\mbox{$m_{_V}$}}
\newcommand{\Mv}{\mbox{$M_{_V}$}}
\newcommand{\peryr}{\mbox{$\>\rm yr^{-1}$}}
\newcommand{\secpoint}{\mbox{$''\mskip-7.6mu.\,$}}
\newcommand{\sqdeg}{\mbox{${\rm deg}^2$}}
\newcommand{\squig}{\sim\!\!}
\newcommand{\subsun}{\mbox{$_{\twelvesy\odot}$}}
\newcommand{\et}{et al.~}

\def\ltsima{$\; \buildrel < \over \sim \;$}
\def\simlt{\lower.5ex\hbox{\ltsima}}
\def\gtsima{$\; \buildrel > \over \sim \;$}
\def\simgt{\lower.5ex\hbox{\gtsima}}
\def\arcs{$''~$}
\def\arcm{$'~$}
\vspace*{0.1cm}
\title{INFRARED OBSERVATIONS OF NEBULAR EMISSION LINES FROM  
GALAXIES at $z \simeq 3$}

\vspace{1cm}
\author{\sc Max Pettini}
\affil{Royal Greenwich Observatory, Madingley Road, Cambridge, CB3 0EZ, UK}
\author{\sc Melinda Kellogg and Charles C. Steidel\altaffilmark{1,2}}
\affil{Palomar Observatory, Caltech 105--24, Pasadena, CA 91125}
\author{\sc Mark Dickinson\altaffilmark{3,4}}
\affil{Department of Physics and Astronomy, The Johns Hopkins University, 
Baltimore, MD 21218}
\author{\sc Kurt L. Adelberger}
\affil{Palomar Observatory, Caltech 105--24, Pasadena, CA 91125}
\author{\sc Mauro Giavalisco\altaffilmark{5}}
\affil{The Carnegie Observatories, 813 Santa Barbara Street, Pasadena, CA 
91101}

\altaffiltext{1}{Alfred P. Sloan Foundation Fellow}
\altaffiltext{2}{NSF Young Investigator}
\altaffiltext{3}{Allan C. Davis Fellow}
\altaffiltext{4}{also Space Telescope Science Institute, 3700 San
Martin Drive, Baltimore, MD 21218}
\altaffiltext{5}{Hubble Fellow}

\newpage
\begin{abstract}
We present the first results from a program of near-infrared spectroscopy
aimed at studying the familiar rest-frame optical 
emission lines from the H~II regions of
Lyman break galaxies at $z \simeq 3$. 
By targeting redshifts which bring the lines of interest into gaps
between the strong OH$^-$ sky emission, we have been 
successful in detecting 
Balmer and [O~III] emission lines in all five 
galaxies observed so far with CGS4 on {\it UKIRT}.
The typical line fluxes are a few times 
$10^{-17}$~erg~s$^{-1}$~cm$^{-2}$, approximately one order of magnitude
lower than the limits reached with wide-field narrow-band imaging surveys.

For a Salpeter IMF and a 
$H_0 = 70$~km~s$^{-1}$~Mpc$^{-1}$, $q_0 = 0.1$
cosmology, 
the \hbeta\ luminosities uncorrected for dust extinction imply
star formation rates of $20 - 270~M_{\odot}$~yr$^{-1}$;
these values are greater than those
which may have been deduced from the ultraviolet continuum 
luminosities at 1500~\AA\ by factors of between 
$\sim 0.7$ and $\sim 7$.
Uncertainties in the shape of the reddening curve and in the 
intrinsic UV continuum 
slope do not allow us yet to assess accurately the level
of dust extinction; however, on the basis 
of the present limited sample it appears that an extinction of 
$1 - 2$ magnitudes at 1500~\AA\ may be typical of Lyman break galaxies.
This value is consistent with recent estimates of dust obscuration
in star forming galaxies at $z \leq 1 $ and does
not require a substantial revision of the broad picture of star 
formation over the Hubble time proposed by Madau et al. (1996).

In four out of five cases 
the velocity dispersion of the emission line gas is 
$\sigma \simeq 70$~\kms, while in the fifth 
the line widths are nearly three times larger.
Virial masses 
$M_{\it vir} \approx 1 - 5 \times 10^{10} M_{\odot}$
are suggested, but both velocities and masses could be higher 
because our observations are only sensitive to 
the brightest cores of these systems 
where the line widths may not sample the full gravitational potential.
The relative redshifts of 
interstellar absorption, nebular emission, and \lya\ emission lines
differ by several hundred \kms\ and suggest that large-scale 
outflows may be a common characteristic
of Lyman break galaxies. The forthcoming availability of high resolution 
infrared spectrographs on large telescopes will soon allow all of these 
questions to be addressed in much greater detail.
\end{abstract}

\newpage
\section{INTRODUCTION}

Star forming galaxies at redshift $z \simeq 3$ are now being  
discovered in large numbers from deep imaging surveys designed to 
reveal the break in their spectral energy distribution
caused by the limit of the Lyman series of neutral hydrogen at 912 \AA\ 
(Steidel, Pettini, \& Hamilton 1995, Steidel et al. 1996; 
Lowenthal et al. 1997). 
At $z = 3$ optical wavelengths sample the rest frame ultraviolet
continuum which, being produced primarily by O and early B type stars, 
is a measure of the instantaneous 
star formation rate.
Thus, by constructing the ultraviolet luminosity function of 
galaxies in different redshift intervals, it has become possible
to sketch the global history of star formation (and associated metal 
production) in the universe
over $\sim 90$\% of the Hubble time (Madau, Pozzetti, \& Dickinson 1998).

In principle, the ultraviolet continuum 
is a more direct avenue to the SFR   
than other commonly used indicators, 
such as the Balmer lines---which measure 
the reprocessed ionizing flux from the most massive stars---and 
the far-infrared luminosity due to grain emission.
Concerns remain, however, regarding the extent to which the 
observed ultraviolet luminosities are attenuated by the dust 
which is likely to be 
associated with the star-forming regions, 
given that even relatively small column densities of dust 
can modify substantially the emergent 
spectral energy distribution in the far-UV.

The magnitude of the dust correction 
appropriate to Lyman break galaxies is 
currently the subject of considerable debate.
On the one hand, there are claims
that the ultraviolet continuum slope and 
UV-optical spectral energy distribution (SED)
imply that the typical $z \simeq 3$ galaxy 
suffers a factor of $\simgt 10$ extinction 
at $\lambda = 1500$~\AA\ 
(Meurer et al. 1997; Sawicki \& Yee 1998),
although other analyses (e.g. Trager et al. 1997)
arrive at more modest estimates (factors of $2 - 6$)
from similar data.
The first results from searches at sub-mm wavelengths
have been interpreted as evidence for a significant
population of dust-obscured star forming galaxies at $z > 2$
(Blain et al. 1998),
but the present
uncertainties in the surface density of sub-mm sources and in their 
redshifts make such conclusions still tentative.
Higher metal production rates than those implied by the 
(uncorrected) ultraviolet luminosity density at high redshift 
have also been advocated to explain the 
lack of evolution in the Fe abundance of rich galaxy clusters up to 
$z \sim 0.3$ (Mushotzky \& Loewenstein 1997; Renzini 1997).

On the other hand, 
dust obscuration does not appear to be a major problem
for galaxies at $z \simlt 1$. 
Observations of the infrared continuum with the {\it Infrared Space Observatory}
(Flores et al. 1998) and of the
\halpha\ emission line
(Tresse \& Maddox 1998; Glazebrook et al. 1998)
indicate that the typical correction to the near-ultraviolet continuum 
luminosity at 2800~\AA\ is at most a factor of two
(see also Treyer et al. 1998).
Similarly, the initial results from a survey of \halpha\ emission 
associated with damped \lya\ galaxies at $z \simeq 2.5$ by Bechtold et al. (1998)
do not point to a significantly higher star formation rate density than
that predicted by Madau et al. (1998).

This confusing situation is unlikely to be settled simply by 
improved measurements of the ultraviolet spectra of high redshift 
galaxies.
As discussed below, uncertainties in the shapes of 
both the unattenuated spectral energy distribution and the 
reddening curve limit the accuracy of even the most careful estimates of 
ultraviolet extinction.
A more profitable approach may be to search for Balmer emission lines, 
particularly H$\alpha$ and H$\beta$,
associated with known galaxies at $z \simeq 3$. 
Since the effects of dust are less severe at rest-frame optical 
wavelengths, a comparison between the star formation rates
implied by the Balmer lines and by the UV continuum
should lead to estimates of dust obscuration which
are less model dependent. 
An added incentive is the possibility of obtaining from
the widths of the nebular 
lines direct estimates 
of the masses of the galaxies which, as argued by Steidel et al. 
(1996), cannot be reliably deduced from 
consideration of the ultraviolet interstellar absorption lines
alone.
Knowledge of the masses associated with Lyman break galaxies 
would make it possible to use their clustering properties to discriminate 
between different cosmologies 
(Adelberger et al. 1998; Steidel et al. 1998b).

With these goals in mind we have begun a pilot program aimed at detecting 
nebular emission lines from galaxies at 
$z \simeq 3$; in this paper we 
report on the initial results of this work.
The brighter objects in our 
spectroscopically confirmed sample 
of Lyman break galaxies are within reach of high
resolution near-infrared spectrographs on 4-m telescopes  
and indeed we have detected Balmer and/or [O~III] emission lines in every 
one of the five galaxies observed so far.
With the forthcoming availability of similar spectrographs on 8-10 m 
telescopes such detections will become routine and may eventually resolve 
the vexed question of dust obscuration at high redshift. \\

\section{OBSERVATIONS AND DATA REDUCTION}

The near-IR sky is dominated by strong and highly variable 
airglow emission from the hydroxyl 
radical (e.g. Ramsay, Mountain \& Geballe 1992)
which normally precludes spectroscopic observations of faint 
sources. At a resolving power $R \simgt 2500$, however,
the OH$^-$ lines are resolved and a significant portion of the spectrum,
particularly in the $K$-band, becomes accessible for work. 
With the large samples of Lyman break galaxies now 
available it is possible to choose objects at favourable
redshifts such that the nebular lines of interest 
fall in gaps between the sky lines.
The CGS4 spectrograph on the UK Infrared Telescope on Mauna Kea, Hawaii
with a $256 \times 256$ InSb array provides both the spectral 
resolution and the wavelength coverage necessary for a search 
for H$\beta$ and 
[O~III] $\lambda\lambda 4959,5007$ at $z \simeq 3$. 

Table 1 gives details of the galaxies observed.
From our Keck LRIS 
spectra, which are reproduced in Figure 1,
we measure {\it two} values of redshift for each galaxy,  
from \lya\ emission (when present) and from the interstellar 
absorption lines; they are listed in columns 4 and 5 respectively
(all redshifts are vacuum heliocentric). 
As discussed below (\S 5),
\lya\ emission is generally redshifted relative to the absorption lines 
by up to several hundred \kms, presumably reflecting large-scale outflows 
in the interstellar media of these galaxies (Kunth et al. 1998). 
Details of the Keck observations and
of the optical and infrared imaging from which the 
$\cal R$ magnitudes and colors in Table 1 were derived
will be presented elsewhere. 
Here we concentrate on {\it UKIRT} observations in the 
$K$-band aimed at detecting \hbeta\ and [O~III] 
emission lines at $z \simeq 3$
and in one case, Q0201$+$113 B13, \halpha\ at 
$z \simeq 2.2$.\footnote{Q0201$+$113 B13 does not qualify as a Lyman
break galaxy according to our usual photometric selection
criteria. It is one of several galaxies in the tail
of the redshift distribution of our sample, 
discovered in the
process of exploring the boundaries for Lyman break galaxies in
the ($U_n - G$) vs. ($G - {\cal R}$) color plane (Steidel et al.
1995).}

The data were obtained over three observing runs in September 1996, and
October and November 1997. 
In 1996 we used the 150 mm focal length camera and 150 grooves~mm$^{-1}$ 
grating (in second order) 
to give a dispersion of 200~\AA\ mm$^{-1}$ (6~\AA\ pixel$^{-1}$); 
with a 1.2~arcsec wide entrance slit and half-pixel stepping of the 
spectrum on the detector to improve the sampling, 
we achieved a resolving power $R \simeq 2600$ corresponding to
$FWHM \simeq 8$~\AA\ in the $K$-band.
In the 1997 observing runs,  
the improved image quality delivered by the new {\it UKIRT} tip-tilt system 
allowed us to use the longer focal length (300 mm) camera;
the corresponding resolution---with the same grating and slit width 
as in 1996---was $FWHM \simeq 6$~\AA.

At the telescope we followed standard observing techniques; a good 
description of the finer points to be taken into account when observing 
faint objects with {\it UKIRT} and CGS4 is given by Eales \& Rawlings (1993).
The galaxies were acquired by blind offsets from nearby stars whose 
relative positions we had previously measured from our own CCD images
with a typical accuracy of $\pm 0.2$~arcsec.
We only used offsets stars within $\approx 2$~arcmin of the galaxies
to make sure that both galaxy and star would fall on the 
dichroic filter used to direct the optical light to the guiding system.
The spectra were recorded on two sets of rows on the detector 
separated by $\sim 20$~arcsec; the object was 
beam-switched between these two positions on the slit
in a ABBA sequence with 600~s 
integration at each position. 

The choice of 600~s as the length of an individual exposure 
(significantly longer than is normally the case when working 
in the infrared) was driven by the need to minimize 
the read-out noise of the detector relative to the 
sky noise in the dark regions of sky between the 
OH$^-$ lines where the Balmer and [O~III] emission lines
are expected to fall.
The penalty which one pays for such long exposures 
is that some of the OH$^-$ lines saturate, or in any case 
vary sufficiently between successive exposures, to make 
it impossible to subtract them out properly. 
In the circumstances, we preferred this course of action (relative to shorter 
exposures), because our observing 
strategy was specifically aimed at {\it avoiding} the
sky emission lines. 

After two sequences of four 600~s exposures
(i.e. approximately every 4800~s) we returned to the offset star and 
realigned the optical and infrared beams (to take into account the
effects of differential atmospheric refraction) before repeating the 
blind offset procedure. 
With total exposure times of between 18\,000 and 
40\,000~s (see Table 1), the final spectra are the sum of many such groups 
of four exposures. Between successive nights we stepped the grating so as 
to shift the spectra by a few pixels 
along the rows of the detector; this reduces the residual fixed pattern 
noise after flat-fielding. In general we summed only data obtained on 
photometric nights with $\simlt 1$~arcsec seeing. One exception are
the October 1997 observations of Q0201$+$113 B13 which were obtained in  
patchy cloud and 1--1.5 arcsec seeing but which were nevertheless found to 
improve the final signal-to-noise ratio 
of the spectrum.

Wavelength calibration was by reference to the 
emission line spectra of Ar and Kr hollow-cathode lamps internal to the 
spectrograph; we further used the sky OH$^-$ emission lines to monitor any 
wavelength shifts during the long series of exposures. 
Observations of standard stars, normally of spectral type A0, provided an 
absolute flux scale for the galaxy spectra.
The major steps of the data reduction procedure are also well described by 
Eales \& Rawlings (1993). 
Briefly, the {\it UKIRT} CGS4 data reduction system provides flat-fielded
two-dimensional images of each group of four exposures 
where the sky background has been subtracted by subtracting the 
sum of the two frames with the object at position  B from the sum of the 
two frames with the object at A.
We used IRAF to wavelength calibrate and rebin these 2-D images to a 
linear scale; we then removed the sky residuals (due mainly to 
temporal variations in the OH$^-$ lines) in each wavelength bin 
by fitting the background along 
the slit with a first order cubic spline.
The last steps involved extracting the galaxy spectra
at positions A and B, multiplying B by $-1$ (since the initial {\it UKIRT} 
processing described above produces a {\it negative} signal at position B),
and adding the resultant two spectra. 
We then co-added the extractions from each group of four exposures to 
produce the final spectra.  
In general we found that using weighted extraction and co-adding algorithms 
made little difference to the final signal-to-noise ratio.\\

\subsection{Results}

Figures 2 and 3 show 
portions of the {\it UKIRT} spectra obtained in 1996 and 
1997 respectively; 
in Figure 4 we have reproduced the inner three cross sections of the 
reduced two dimensional images from the 1996 observations.
In Table 2 we list the measured redshifts and fluxes
of the nebular emission lines covered, together with the line luminosities 
deduced for $H_0 = 70$~km~s$^{-1}$~Mpc$^{-1}$ and $q_0 = 0.1$ (we adopt 
these values throughout the paper).  
From the continuum fluxes implied by the $K_{\rm AB}$
magnitudes (see Table 1) 
we calculate the values of rest frame equivalent width
listed in Table 2.
It can be seen from the figures that, 
by targeting redshifts such that the 
nebular lines are well removed from the strongest night sky features,
we have generally succeeded in detecting [O~III] and Balmer emission lines from 
the H~II regions of Lyman break 
galaxies at $z \simeq 3$. 
Even so, it is clear that
the observations are at the limit 
of what can be achieved with 4-m class telescopes,
resulting in detections at levels of only a few $\sigma$ 
(see Table 2) despite the very long exposure times.
In general [O~III]$\lambda 4959$ is 
in the noise, but in each case the upper limit on its flux 
is consistent with the measured flux of $\lambda 5007$ which 
is three times stronger. In one case, Q0201$+$113 C6, we do detect 
[O~III]$\lambda 4959$ but not $\lambda 5007$ because 
at $z_{\rm em} = 3.094$ the latter coincides with a strong OH$^-$
line at $2.04993~\mu$m which is saturated in our spectra. 
In Q0000$-$263 D6, \hbeta\ falls outside the $K$-band window.

Typical line fluxes are a few times
$10^{-17}$~erg~s$^{-1}$~cm$^{-2}$, approximately one order of
magnitude lower than the limits achieved so far in wide field
surveys in the near infrared using narrow band imaging (e.g.
Thompson, Mannucci, \& Beckwith 1996). This is likely to be the
reason why such surveys have not yet found a widespread 
population of star-forming galaxies at high redshift.
Targeted searches for \halpha\ emission at $z \simeq 2$
in QSO fields, using narrow band filters tuned to
the redshifts of known absorbers,
have been more successful (Teplitz, Malkan, 
\& McLean 1998; Mannucci et al. 1998; Bechtold et al. 1998).\\

\section{STAR FORMATION RATES AND DUST EXTINCTION}

In Table 3 and Figure 5 we compare the values of the star 
formation rate (SFR) 
deduced in the five galaxies from the measured luminosities in the 
\hbeta\ line
and in the far-UV continuum ($\lambda_0 = 1500$~\AA) respectively. At 
this stage we do {\it not} correct the observed fluxes 
for dust obscuration.
In calculating SFR$_{{\rm H}\beta}$ we 
have assumed a ratio \halpha/\hbeta = 2.75 
(Osterbrock 1989) and Kennicutt's (1983) calibration 
SFR = $L_{{\rm H}\alpha}/1.12 \times 10^{41}~M_{\odot}$~yr$^{-1}$ 
(where $L_{{\rm H}\alpha}$ is in erg~s$^{-1}$) which is appropriate for a 
Salpeter IMF from $M = 100$ to $0.1~M_{\odot}$.\footnote{Madau et
al. 1998 deduced a slightly different conversion 
factor, 
SFR = $L_{{\rm H}\alpha}/1.58 \times 10^{41}~M_{\odot}$~yr$^{-1}$,
using more recent population synthesis models. 
Adopting this calibration would {\it decrease} the values
of SFR$_{\rm H\beta}$ in column 3 of Table 3 by 29\%\,.}
For Q0000$-$263 D6, where only [O~III]$\lambda 5007$ is observed,
we derive an upper limit to SFR$_{{\rm H}\beta}$
assuming that $\lambda 5007$/H$_{\beta} \geq 2$.
This is a conservative limit based on data from 
local star-forming regions (Terlevich et al. 1991; 
Stasinska \& Leitherer 1996), 
as well as the three Lyman break galaxies in the present 
sample where both [O~III] and \hbeta\ are detected (see Table 2). 

Turning to the UV luminosity, it is important to realize that, 
although the continuum at 1500~\AA\ is dominated 
by short-lived O and B stars, it does nevertheless show 
a modest increase with time 
in {\it continuous} star formation models.
(In the case of an instantaneous burst of star formation 
the UV luminosity obviously fades 
with time after the starburst).\footnote{Throughout this
paper we only consider `continuous' star formation models---in which gas 
is converted into stars at a constant rate, as opposed to 
`instantaneous burst' models---in which all stars are assumed to be formed 
at time $t = 0$ with negligible star formation thereafter.
Clearly these are two limiting cases to the real star formation 
history of Lyman break galaxies. 
However, in our view single burst models are less likely to 
apply to the galaxies observed here because the Balmer 
and [O~III] emission lines would fade rapidly after such a burst 
and would be undetectable at 
the sensitivity of the present observations 
after less than $10^7$ years (Stasinska \& Leitherer 1996).}
The Bruzual \& Charlot (1996) models predict 
that, for a Salpeter IMF within the above limits and solar metallicity,
the UV luminosity at 1500 \AA\ from a region undergoing continuous 
star formation tends to an asymptote at 
$10^9$~years, at which point a ${\rm SFR} = 1 M_{\odot}$~yr$^{-1}$
produces $L_{1500} = 10^{28}$~erg~s$^{-1}$~Hz$^{-1}$.
For comparison, $10^7$~years after the onset of star formation
the UV luminosity corresponding to ${\rm SFR} = 1 M_{\odot}$~yr$^{-1}$
is only $\sim 60$\% of the above value 
(see also Leitherer, Robert, \& Heckman 1995).
In the Bruzual \& Charlot models these conversion factors 
do not depend sensitively on  metallicity.\footnote{An empirical 
determination of the metallicity dependence of the integrated UV 
continuum of stellar populations
awaits the availability of the appropriate stellar libraries.}

At $z \simeq 3$ our $\cal R$ filter 
(with effective wavelength $\lambda = 6850$~\AA)
samples the rest-frame continuum  
near $\lambda_0 = 1710$~\AA.
Since our magnitudes are on the AB scale
(${\rm AB} = -48.60 - 2.5 {\rm log} f_{\nu}$),
the median ${\cal R} = 23.5$ of the Lyman break
galaxies considered here
corresponds to 
$f_{\nu} = 1.5 \times 10^{-29}$~erg~s$^{-1}$~cm$^{-2}$~Hz$^{-1}$;
at $z = 3$ this in turn translates to a continuum luminosity
$L_{1710} = 3.1 \times 10^{29}$~erg~s$^{-1}$~Hz$^{-1}$\,.
In arriving at the star formation rates listed in columns (4) and (5) of 
Table 3, we have applied small $k$-corrections
(generally less than 10\%), based on the  
spectral slopes measured from our Keck LRIS spectra, 
to deduce $L_{1500}$ from the values of $L_{1710}$ implied by the 
galaxy ${\cal R}$ magnitudes. 

There are two conclusions that can be drawn from the results in 
Table 3. First, in most cases the Balmer lines indicate star formation 
rates which broadly agree, to within factors of $2-3$, with the values
obtained by applying the Bruzual \& Charlot (1996) models
to the observed UV continuum luminosities. 
This is the level of agreement 
to be expected between different star formation indicators,
even in the local universe (e.g. Meurer et al. 1995), 
given the systematic uncertainties in the 
calibrations and the possibility that
ionized gas and early-type stars have 
different spatial distributions (e.g. Leitherer et al. 1996).

Second, Figure 5 suggests that
there may be a trend between the ($G - {\cal R}$) 
color and SFR$_{{\rm H}\beta}$/SFR$_{\rm UV}$ in 
the sense that the redder galaxies have apparently the larger values
of this ratio. Thus Q0201$+$113 B13, which is among the bluest galaxies in our 
entire sample, has a Balmer line flux close to that predicted from 
its UV continuum luminosity,  
while in DSF~2237$+$116~C2, which with ($G - {\cal R}$) = 1.13 is close to the 
red limit of our selection criteria for Lyman break galaxies,
we apparently see $4 - 7$ times more \hbeta\ photons than expected.
The other three galaxies are intermediate cases.
Before reading too much into this `trend', we must remember that
we are dealing with only a small number of measurements from 
infrared spectra of low signal-to-noise ratio. 
Nevertheless, the effect is in the direction expected from dust 
extinction, which is more effective at ultraviolet wavelengths than in 
the optical. It is worthwhile considering, then,
the magnitude of the UV extinction at 1500~\AA, $A_{1500}$, implied by
the SFR$_{{\rm H}\beta}$/SFR$_{\rm UV}$ ratios, and how this is related to the 
observed ($G - {\cal R}$) colors. \\

\subsection{Ultraviolet Extinction of Lyman Break Galaxies}

Specifically, we are going to determine the values of 
$A_{1500}$ which satisfy the requirement 
${\rm SFR}^{\prime}_{\rm UV} = {\rm SFR}^{\prime}_{{\rm H}\beta}$,
where ${\rm SFR}^{\prime}$ are the star formation rates corrected
for extinction.
This would be a straightforward calculation were it not for the fact 
that we are totally ignorant of the wavelength dependence 
of dust extinction in Lyman break galaxies! 
Here we consider what are generally thought to be 
two limiting cases, the extinction curve determined for stars 
in the Small Magellanic Cloud 
(Bouchet et al. 1985; Gordon \& Clayton 1998)
and the `attenuation' curve 
derived from the integrated spectra of local star forming galaxies 
(Calzetti, Kinney, \& Storchi-Bergmann 1994; Calzetti 1997a). 
The former rises steeply towards shorter wavelengths, 
while the latter is `greyer' (see for example Figure 
2a of Calzetti 1997a). These differences 
are thought to arise, at least in part, from different 
geometrical configurations of the dust and background sources of light. 
The SMC extinction curve is derived from observations of individual stars,
where dust is mostly in a foreground screen which removes both scattering and 
absorption components of the extinction curve, 
whereas the integrated stellar spectrum which 
emerges from an extended region
presumably includes photons scattered {\it into} the line of sight by 
dust mixed with the stars.
Whether the latter situation also applies to 
Lyman break galaxies depends to some extent 
on the geometrical configuration of the 
large scale outflows which seem to be a common feature of their 
interstellar media (see \S 5 below).

For the Bouchet et al. (1995) SMC reddening law with Pei's (1992) 
normalisation
\begin{equation}
	A_{1500} = 3.44 \times {\rm log~(SFR}_{{\rm H}\beta}/{\rm SFR_{\rm UV})}
	\label{}
\end{equation}¥
where $A_{1500}$ is in magnitudes, whereas
the Calzetti (1997a) attenuation curve gives
\begin{equation}
	A_{1500} = 4.87 \times {\rm log~(SFR}_{{\rm H}\beta}/{\rm SFR_{\rm UV})}.
	\label{}
\end{equation}¥
With these conversion factors we deduce the values of $A_{1500}$
listed in Table 4. It can be readily seen that the uncertainties in the 
conversion from $L_{1500}$ to SFR and in the wavelength dependence of the 
extinction combine to give a wide range of possibilities for 
$A_{1500}$. UV extinctions by $\approx 1 - 2$ magnitudes
seem typical, but we also find some unlikely solutions. 
The large dust corrections indicated for DSF~2237$+$116~C2 if the Calzetti 
reddening curve applies would imply 
star formation rates in excess of 1000~$M_{\odot}$~yr$^{-1}$\,.
We have retained negative (and therefore unphysical)
values of $A_{1500}$ in Table 4 because they 
give an indication of the inherent limitations 
of the analysis, which forces 
${\rm SFR}^{\prime}_{\rm UV} = {\rm SFR}^{\prime}_{{\rm H}\beta}$.
Other possibilities are that our assumption that
$\lambda 5007$/H$_{\beta} \geq 2$ in Q0000$-$263 D6 is incorrect,
and that in Q0201$+$113 B13 the Balmer lines suffer higher extinction
than the stellar continuum (or some of the ionizing photons escape the 
nebula). 
Note that the combination of the $10^9$~year old 
continuous star formation model, which 
gives the lower SFR for a given UV luminosity,
and the Calzetti attenuation, which has the lower differential extinction 
between 1500~\AA\ and \hbeta, is the one which 
produces the largest values of $A_{1500}$.

\subsubsection {Extinction Estimates from the Slope of the Ultraviolet 
Continuum}

With a larger sample of emission line measurements, it will be possible in 
future to obtain an independent estimate of the star formation rate 
density at 
high $z$ which is less sensitive to dust corrections than the 
current rest frame UV data.
For moment, however, it is of interest to consider whether the few
Balmer line detections available can be used to `calibrate', as it were,
a dust extinction index based on the ultraviolet spectral energy 
distribution (SED), given the large 
number of UV spectra of Lyman break galaxies 
now being obtained (Steidel et al. 1998b).

An index often used is $\beta$, the slope of the UV continuum approximated 
by a power law of the form $F_{\lambda} \propto \lambda^{\beta}$.
Model spectral energy distributions (e.g. Leitherer \& Heckman 1995)
show that $\beta$ changes relatively little with age and metallicity; 
for a Salpeter IMF from $M = 100$ to $0.1~M_{\odot}$
the Bruzual \& Charlot (1996) continuous star formation models have 
$\beta$ increasing from $\sim -2.5$ to $\sim -2$
over the period $10^7$ to $10^9$ years.
Empirically, the measured values of $\beta$ 
correlate both with the 
far-infrared excess (Meurer et al. 1995)
and with the reddening deduced from the Balmer decrement 
(Calzetti 1997a), 
although there is considerable scatter in these relations.
The empirical starburst template spectrum derived by Calzetti (1997b)
has $\beta = -2.1$ which would suggest that the youngest model SEDs may be 
too blue.

At $z \simeq 3$ our ($G - {\cal R}$) color provides a more 
reliable estimate of the continuum slope between 1190 and 1710~\AA\ 
(the effective wavelengths 
of the two filters are 4740 and 6850 \AA\ respectively) 
than the value which could be measured directly from the Keck spectra.
The reason for this is that the spectra are obtained through 
multi-object apertures which have fixed orientation on the sky; 
differential atmospheric refraction can lead to increasing light losses 
with decreasing wavelength which may be difficult to calibrate out.
The net effect is to artificially redden the spectra (by an unknown amount)
and it is therefore dangerous to use published Keck LRIS spectra 
to measure $\beta$. 
For comparison with other analyses,
we have listed in column (3) of Table 5 
the values of $\beta$ implied by the observed 
($G - {\cal R}$) colour for each of the five
galaxies in the present sample. 
They were calculated by convolving power 
laws in $F_{\lambda}$ with the $G$ and  ${\cal R}$
filter transmission curves and 
folding in the average
Lyman~$\alpha$ forest opacity as a function of redshift 
according to the prescription 
by Madau (1995).\footnote{Along any particular sight-line the 
\lya\ forest opacity may differ somewhat from Madau's average. 
However, the ensuing error in slope of the 
UV continuum is small compared with that which could result
from the uncertain photometric calibration of the LRIS spectra.}

In order to estimate the dust reddening of the UV continuum, we have 
determined a color excess 
E($G - {\cal R}$) = ($G - {\cal R}$)$_{\rm obs}$ $-$ 
($G - {\cal R}$)$_{\rm calc}$
where the two terms on the right-hand side of the equality 
are the observed and expected 
values of ($G - {\cal R}$).
Again, we have calculated ($G - {\cal R}$)$_{\rm calc}$ 
by convolving the appropriately redshifted
Bruzual \& Charlot (1996) model SEDs with the 
filter transmission curves and allowing for the 
\lya\ forest opacity.
The resulting values of E($G - {\cal R}$) for the  
$10^7$ and $10^9$ year model SEDs are listed 
in columns (4) and (7) of Table 5 respectively.
Columns (5) and (6), and (8) and (9) 
of Table 5 then give the values of $A_{1500}$
implied by the ($G - {\cal R}$) color excess, depending on whether the 
SMC or Calzetti extinction law is adopted. Once again we see that
$A_{1500}$ is not well constrained, the different permutations
of intrinsic UV slope and reddening curve allowing solutions which span a 
$1 - 1.5$ magnitude range. 
In this case it is the combination of the Calzetti curve and 
$10^7$ year model (with its very blue intrinsic 
slope) which requires the largest values of ultraviolet extinction.

We can proceed further, however, by considering which set of model assumptions
results in the values of $A_{1500}$ for each galaxy which are most  
consistent between Tables 4 and 5.
Comparing the entries in the two tables we see that for 
Q0000$-$263 D6 the combination of $10^9$~years SED (UV9) and SMC extinction law
satisfies both the UV slope and the 
SFR$_{{\rm H}\beta}$/SFR$_{\rm UV}$ tests with
$A_{1500} \simeq 0.5$\,.
UV7 + SMC gives a good solution in Q0201$+$113 C6 for
$A_{1500} \simeq 0.7$.
The agreement is less than optimum in B2~0902$+$343~C6 and DSF~2237$+$116~C2, but 
even here we see that the UV7 SED gives answers which differ by only 
$\approx 0.5$~magnitudes between Tables 4 and 5 independently of the 
extinction curve used.
Q0201$+$113 B13 is not useful in this context because its spectrum is 
essentially unreddened.

It would be inappropriate to take these comparisons too literally and, 
for example, conclude that a particular extinction law is favoured
or deduce an age for the star forming galaxies. The reasons are that
our infrared spectra are of low signal-to-noise ratio and 
the galaxies probed are not sufficiently reddened to  
discriminate positively between the models.  
Furthermore, we do not even know at this stage that the models 
are internally consistent. For instance, the 
Bruzual \& Charlot (1996) SED for the $10^7$ year old continuous star 
formation case may have a UV slope which is too blue compared to the 
unreddened continuum of the typical Lyman break galaxy and yet provide the 
correct calibration of the luminosity at 1500~\AA\ as a 
function of SFR---the slope and normalisation of the models 
are independent parameters to some extent. 

What seems to emerge from the comparison of Tables 4 and 5, however, 
is that the values of $A_{1500}$ which fit best both the UV 
slopes and the ratios of the Balmer lines to the UV luminosity are 
in every case 
{\it intermediate} between the extreme possibilities 
allowed by the different models. Very high and very low 
values of UV extinction seem to be excluded even by the present limited 
sample.\\

\section{LINE WIDTHS AND MASSES OF LYMAN BREAK GALAXIES}

In column (2) of Table 6 we have listed the values of the velocity 
dispersion $\sigma$~(=~FWHM/2.355)
deduced by fitting the Balmer and [O~III] emission  
lines with Gaussian profiles and correcting for the instrumental resolution
(thermal motions make a very small contribution to the line widths
found here). 
In general there is good agreement between values of $\sigma$  
measured from different emission lines in the same galaxy; 
the errors quoted in Table 6 were obtained by propagating the error
of each Gaussian fit. We find that four out of the five galaxies have similar 
velocity dispersions, $\sigma \simeq 60 - 80$~km~s$^{-1}$, 
while in the fifth, DSF~2237$+$116~C2, the lines are apparently 
about three times wider. 

{\it HST} NICMOS images of two of the galaxies in our 
sample (see Table 6) show well resolved objects with half-light radii 
$r \sim 0.25$~arcsec (corresponding to $\approx 2$~kpc at 
$z \simeq 3$ in the cosmology adopted here).
For two other galaxies, we measure similar sizes 
from near-IR images obtained in 0.5~arcsec seeing 
with NIRC on the Keck~II telescope
(when both are available, 
ground-based and {\it HST} measurements are in good agreement).
Combining these radii with the measured velocity dispersions we obtain 
virial masses $M_{\it vir} \approx 1 - 5 \times 10^{10}~M_{\odot}$
(see column (5) of Table 6).
Thus, in Lyman break galaxies at 
$z \simeq 3$ we already find masses comparable to, or exceeding, 
that of the Milky Way bulge (Dwek et al. 1995).
In reality, we may well be 
underestimating the total masses of these systems because
at the low S/N of our observations we could easily  
miss broader components of the emission 
lines\footnote{It is also conceivable that we 
may {\it over}-estimating the virial masses
if the emission line widths have significant contributions from gas
accelerated to high velocities from supernova explosions and stellar 
winds.}.
As discussed by Steidel et al. (1998a --- see also Giavalisco et al.
1998 and Adelberger et al. 1998)
the clustering properties of Lyman break galaxies, interpreted
within the cold dark matter scenario of galaxy formation, point to dark  
halo masses $M_{\rm DM} \simgt 10^{11}~M_{\odot}$.

If star formation takes place in rotationally supported disks, 
the typical $\sigma \simeq 70$\kms\ implies 
rotational velocities $v_{\rm rot} \simeq 120$\kms\ (Rix et al. 1997).
However, it has been known for a long time
that in local starburst galaxies 
the nuclear emission lines do not reflect 
the full rotation speed of the 
galaxy (Weedman 1983).
Lehnert \& Heckman (1996) found that starbursts 
are roughly confined to solid body part of the galaxy rotation curve
and that the widths of the emission lines typically underestimate  
the full $v_{\rm rot}$ by a factor of $\approx 2$.
If a similar situation applied to galaxies at $z \simeq 3$
(a highly speculative hypothesis, admittedly)
rotational velocities $v_{\rm rot} \approx 200 - 250$\kms\ 
may be indicated.
Note that in this picture, where the emission in the 
nebular lines is dominated by H~II regions in the star forming 
cores of larger galaxies,
it is reasonable to expect the strong, saturated
interstellar absorption lines to be broader than 
\hbeta\ and [O~III], as is indeed the case (Steidel et al. 1996),  
because the absorption takes place over longer 
pathlengths, presumably through half of the galaxy.
In this case {\it both} galactic rotation and the large scale `stirring'
of the interstellar gas by  
supernovae and stellar winds (see \S 5 below)
may be contributing to the strengths of the 
absorption lines. Heckman (1998) reached a similar 
conclusion for local starburst galaxies. \\

\section{REDSHIFTS OF EMISSION LINES AND LARGE SCALE MOTIONS}

Table 7 lists the relative velocities of the interstellar absorption,
nebular emission, and \lya\ emission lines;
we have {\it assumed} that
the Balmer and [O~III] emission lines are at the systemic redshift
of each galaxy.
The immediate conclusion is that large velocity fields are 
a common feature of Lyman break galaxies, confirming the initial hints
provided by the large equivalent widths of
the interstellar absorption lines (Steidel et al. 1996) and
in agreement with the analyses by Lowenthal et al. (1997) and Franx et 
al. (1997).
In all three cases where we detect \lya\ emission, 
the line peak is {\it redshifted} by $\approx 1000$~\kms\ 
relative to the metal absorption lines. 
In two out of three cases
(Q0000$-$263 D6 and  B2~0902$+$343~C6) \hbeta\ and [O~III] are at intermediate 
velocities; furthermore, in Q0000$-$263 D6  
\lya\ emission exhibits
an obvious P-Cygni profile, with a sharp drop on the blue side and a long 
tail of emission extending to $\sim +1100$~\kms\ on the relative velocity 
scale of Table 7 
(see Figure 8 of Pettini et al. 1997).

These characteristics are most easily interpreted as evidence for large 
scale outflows with velocities of $\simgt 500$~\kms\ in the interstellar 
media of the galaxies observed. In this picture
\lya\ emission is suppressed by 
resonant scattering and the only \lya\ photons which can escape 
unabsorbed in our direction are those 
back-scattered from the far side of the expanding 
nebula, whereas in absorption against the stellar continuum
we see the approaching part of the outflow.
The data also show that the real situation is probably more complex 
than this simple sketch.
In one case (DSF~2237$+$116~C2) \hbeta\ and [O~III] emission are apparently at 
roughly the same velocity as the absorption lines, even though 
\lya\ emission is redshifted by $\approx 1000$~\kms. 
Most difficult to understand in the
$3200$\kms\ difference between emission and absorption lines 
in Q0201$+$113 C6, large enough to raise the question of whether the two sets of 
lines are at all related!

Nevertheless, all of the above features---blueshifted metal lines, 
redshifted \lya\ emission, and a variety of \lya\ profiles ranging
from emission 
to P-Cygni and to damped absorption---have been observed in {\it HST}
spectra of nearby H~II and starburst galaxies  
(Kunth et al 1998; Gonz\'{a}lez Delgado et al. 1998), 
although lower outflows velocities are normally involved.
The variety of \lya\ emission-absorption profiles may 
reflect different stages in the interaction of the mechanical energy 
generated by the starburst with the interstellar medium of the host galaxy
and/or different viewing angles 
(Giavalisco, Koratkar, \& Calzetti 1996).
Irrespective of the details of this picture, two facts are clear.
First, we see {\it directly} the process by which heavy elements can be 
distributed far from the sites of production (Heckman 1998).
If outflows of $\sim 500$~\kms\ are maintained over lifetimes of 
$\simgt 1 \times 10^8$~years, 
the metals will travel over distances 
$\simgt 50$~kpc, more than enough to seed 
the entire dark matter halo of each Lyman break galaxy.
Second, just as the luminosity of \lya\ emission is in general poorly
related to the star formation rate, its wavelength is not a useful measure 
of the galaxy systemic redshift.\\

\section{DISCUSSION}

Our first detections of nebular emission lines at $z \simeq 3$ 
have predictably gone only part way towards
determining the typical extinction 
suffered by the ultraviolet continuum of Lyman break galaxies.
The present sample of 5 galaxies includes one which is essentially 
unreddened, three where the continuum at 1500~\AA\ is dimmed by between 
$\sim 0.5$ and $\sim 1$ magnitude, and one where the extinction may be as 
much as $\sim 2.5$~magnitudes. 
The main conclusion reached in \S 3, however,  
is that the values of $A_{1500}$ which best fit all the available 
data are intermediate solutions within the wide range allowed by 
different combinations of SEDs and reddening curves.

Recently, Dickinson (1998) has analysed the ($G - {\cal R}$) colors 
of our entire sample of 
more than 400 spectroscopically 
confirmed Lyman break galaxies
in the same way as discussed in \S 3.1.1 above.
His analysis showed that, based on the UV slope alone, 
the value of $A_{1500}$ to be applied to the population as a
whole
can be as little as 0.75 magnitudes or as large as 3.1 magnitudes
for the different combinations of SEDs and extinction curves considered in
\S 3.1.1\,.
The upper end of this range
agrees with the factor of 16 
proposed by Sawicki \& Yee (1998); in both cases
such high values are derived assuming the bluest intrinsic continuum
(UV7 in the notation above) and the greyest extinction curve
(Calzetti). 

If our preliminary conclusion for the five galaxies considered here
applies to the whole population, it would
favour values of $A_{1500}$ near the middle of the range 
determined by Dickinson; thus, 
$A_{1500} \simeq 1 - 2$ may be the typical 
ultraviolet extinction (in magnitudes) suffered by Lyman break  
galaxies at 1500~\AA. If this still tentative statement holds up in the 
light of future infrared observations, several interesting 
consequences follow.

First, this amount of dust would produce an extinction
$A_{2800} \simeq 0.5 - 1.5$~magnitudes (SMC and Calzetti curves 
respectively) of the {\it near}-ultraviolet continuum 
of galaxies at $z \simlt 1$ in the {\it Canada-France Redshift Survey}
(Lilly et al. 1996).
This is in good agreement with  available determinations of
$A_{2800}$, based on the
comparison between the luminosities in \halpha\ and in the 
continuum at 2800~\AA\ (Tresse \& Maddox 1998; Glazebrook 1998; 
see also Flores et al. 1998).
Similarly, Buat \& Burgarella (1998) recently concluded that  
$A_{2000} \simeq 1.2$~magnitudes
is typical of nearby starburst galaxies. 
Thus there is no need to assume that, as we look back to earlier epochs, 
an increasing fraction of the star formation activity takes place in highly 
obscured galaxies (such a scenario would be called for 
if the far-UV continuum of Lyman break galaxies were 
suppressed by more than a factor of 10; Madau et al. 1998, 
Guiderdoni et al. 1997).
The dust corrections proposed here, while raising the values of 
the ultraviolet luminosity density by larger factors at $z > 3$ than 
at $z < 1$, still maintain the same broad picture of the cosmic star 
formation history sketched by Madau et al. (1996), 
with the peak in activity at an intermediate epoch between these two 
redshifts. 

(Of course the present data do not address the separate question of whether 
our photometric selection based on the Lyman break misses a significant 
component of the galaxy population at $z \simeq 3$, in which the 
the UV continuum may be too faint to be detected.
However, as discussed
by Adelberger et al. (1998), the
clustering properties of the Lyman break galaxies
suggest that such an unseen population is likely to contribute only 
a small fraction of the total number of {\it luminous} galaxies
at these redshifts.)

Second, it is of interest to ask where the metals associated with the
higher, dust-corrected rates of star formation are to be found.
Pettini et al. (1997) showed that at $z \simgt 2$
damped \lya\ systems can account for 
the metal production expected from the {\it uncorrected}
UV luminosity. 
We note, however, that the estimates of metallicity by Pettini et al. 
are based on measurements of the abundance of Zn, 
which traces that of the Fe-peak elements.
The average metallicity of DLAs, $Z_{DLA}$, may need to be corrected 
upwards by a factor of $\approx 2$, if O and other $\alpha$ elements 
are overabundant by 0.4 dex relative to Fe, as is the case in 
metal-poor stars of the Milky Way (e.g. McWilliam 1997).
Therefore even when dust obscuration is taken 
into account, there may still be a broad consistency, given the 
uncertainties, between these two 
independent measures of the metallicity of the 
universe---the metals seen in absorption towards distant QSOs and those
produced by the star forming galaxies imaged directly.
On the other hand,
if future observations were to reveal 
that we have underestimated the dust 
correction to the UV luminosity of $z \simeq 3$ galaxies,
the implication would be that DLA systems 
`miss' metal-rich regions of the universe,
either because current QSO absorption line samples are biased in favour
of metal- and dust-poor sightlines, and/or because the metals may be retained 
close to their production sites. 
It will be possible to test these ideas in the near future, as 
wide field surveys yield
new and larger samples of damped \lya\ systems 
towards fainter---and potentially more 
reddened---QSOs 
than studied up to now (e.g. Shaver et al. 1998).\\

\section{SUMMARY}

The main results of this work are as follows.

(1) We report the first detections of Balmer and [O~III] 
emission lines from five Lyman break galaxies at $z \simeq 3$, 
achieved by targeting redshifts such that these transitions 
fall in the gaps between 
the strong OH$^-$ lines which dominate the near-infrared sky. 
The five galaxies span nearly the full range of our color 
selection criteria for Lyman break galaxies.

(2) For a Salpeter IMF between 100 and 0.1 $M_{\odot}$
and neglecting dust extinction, the 
measured \hbeta\ luminosities of $\sim 0.8 - 10 \times 
10^{42}$~erg~s$^{-1}$ ($H_0 = 70$~km~s$^{-1}$~Mpc$^{-1}$,
$q_0 = 0.1$) imply star formation rates of between $\sim 20$ and
$\sim 270$~$M_{\odot}$~yr$^{-1}$. These values are 
generally larger than those estimated from the ultraviolet continuum at 
1500~\AA\ by factors of between $\sim 0.7$ and $\sim 7$. 

(3) The present sample is too small to determine reliably 
the value of $A_{1500}$, the dust extinction suffered by the continuum at 
1500~\AA,  given the 
uncertainties in the model spectral energy distributions and in the 
wavelength dependence of the extinction curve.
However, in general we find that 
the solutions which best fit {\it both} the slope and the luminosity of 
the UV continuum (relative to \hbeta)
correspond to intermediate values of extinction within the large range 
allowed  by the models.
Thus, we favour $A_{1500} \simeq 1 - 2$ magnitudes as the values
most likely to be representative of the whole sample 
of Lyman break galaxies.

(4) The amount of dust implied would depress 
the {\it near}-UV continuum by only $A_{2800} \simeq 0.5 - 1.5$~magnitudes,
in good agreement with recent 
determinations of this quantity for {\it CFRS} galaxies 
at $z < 1$. 
The upward corrections proposed do not require a significant revision 
of the broad picture of the star formation history of the universe 
put together by Madau et al. (1996). 
Similarly, there is still a plausible agreement between the 
metal production associated with the star forming galaxies we see and 
the heavy elements detected in absorption against distant QSOs.

(5) The nebular emission lines recorded are resolved in every case.
It appears that the 
typical velocity dispersion is $\sigma \simeq 70$~\kms, although 
we also find one galaxy where the line widths are
nearly three times larger.
With typical half-light radii of $\sim 2$~kpc, virial masses 
$M_{\it vir} \approx 1 - 5 \times 10^{10}~M_{\odot}$ 
are suggested.
It is possible that both velocities and masses have been
underestimated because with the limited sensitivity
of the present observations 
we may well be seeing only 
the inner cores of the galaxies, where star formation is most in
evidence.

(6) The relative redshifts of the interstellar absorption, nebular 
emission, and \lya\ emission lines indicate that large scale 
outflows with velocities of at least $\approx 500$~\kms\ are a common 
feature of the interstellar media of Lyman break galaxies.
These outflows, presumably driven by the mechanical energy generated in 
the star formation episodes, can distribute the products of stellar 
nucleosynthesis over large volumes.

(7) Perhaps the most valuable aspect of this work is in showing that
infrared spectroscopy of galaxies at $z \simeq 3$ is feasible 
and in highlighting
the wealth of information which such spectra potentially offer. 
With the forthcoming availability of medium resolution spectrographs
on large telescopes, we expect that infrared observations will
play a major role in advancing our understanding 
of the nature of Lyman break galaxies.\\

\acknowledgements

We are grateful to the {\it UKIRT} time assignment committee for their 
continuing support of this work, and to the staff 
at {\it UKIRT} and at the Joint Astronomy Centre 
in Hilo for their generous and competent assistance with the observations.
The interpretation of these results has benefited much 
from discussions with many colleagues, in particular Daniela Calzetti, 
Karl Glazebrook, 
Tim Heckman, Claus Leitherer, Piero Madau, Gerhardt Meurer, and Roberto 
Terlevich. 
We are grateful to the referee, Arjun Dey, for constructive comments
which improved the paper.
C. C. S. acknowledges support from the National Science Foundation through 
grant AST~94-57446 and from the Alfred P. Sloan Foundation.
M. G.  has been supported through grant HF-01071.01-94A from the Space 
Telescope Science Institute, which is operated by the Association
of Universities for Research in Astronomy, Inc., under NASA contract 
NAS~5-26555.

\hspace*{1cm}
\begin{figure}
\vspace*{-1.5cm}
\figurenum{0}
\epsscale{1.1}
\plotone{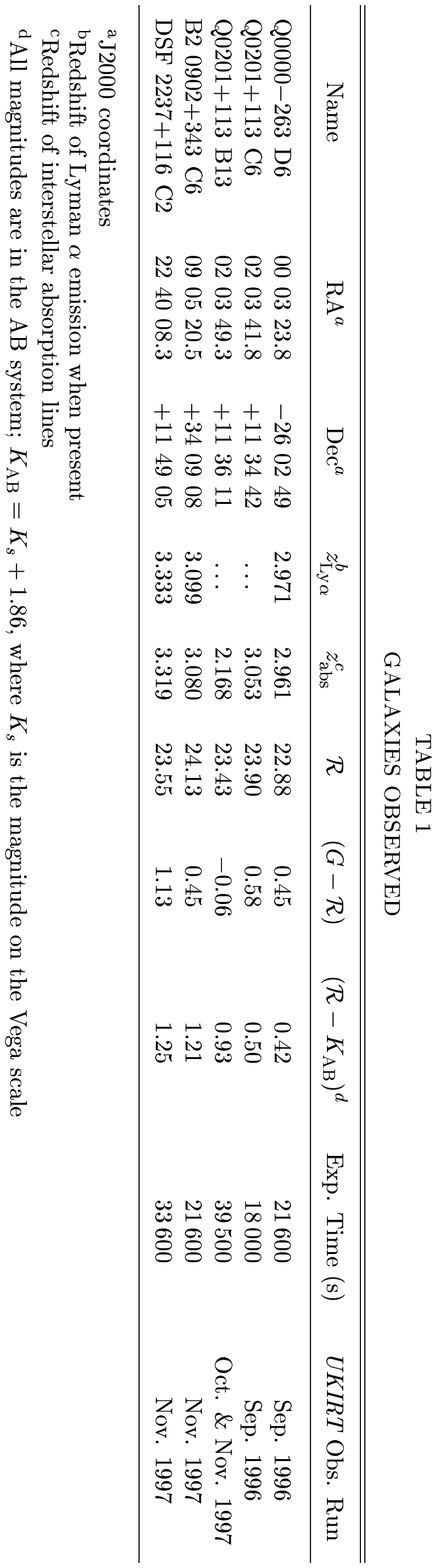}
\end{figure}

%

\begin{figure}
\figurenum{0}
\epsscale{1.0}
\plotone{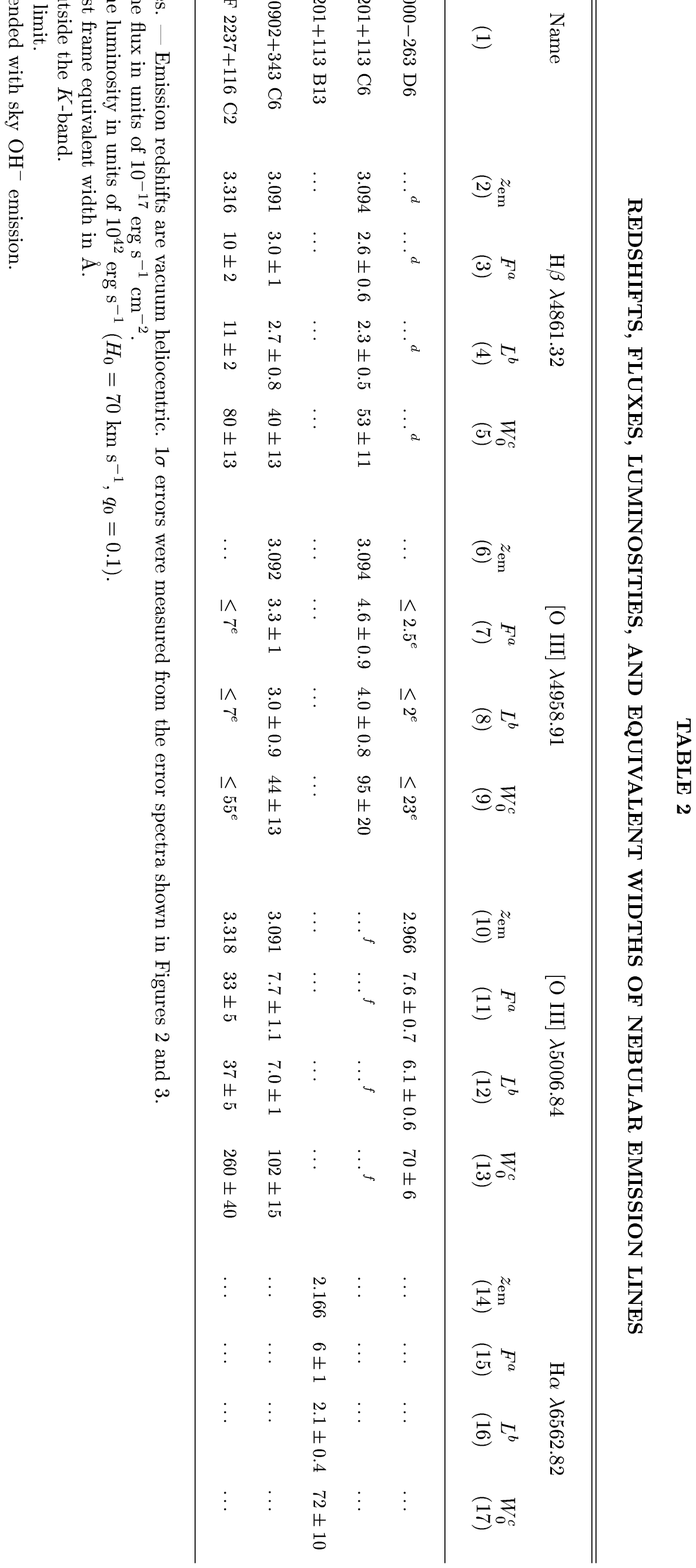}
\end{figure}


\hspace*{1cm}
\begin{figure}
\vspace*{-1.5cm}
\figurenum{0}
\epsscale{1.2}
\plotone{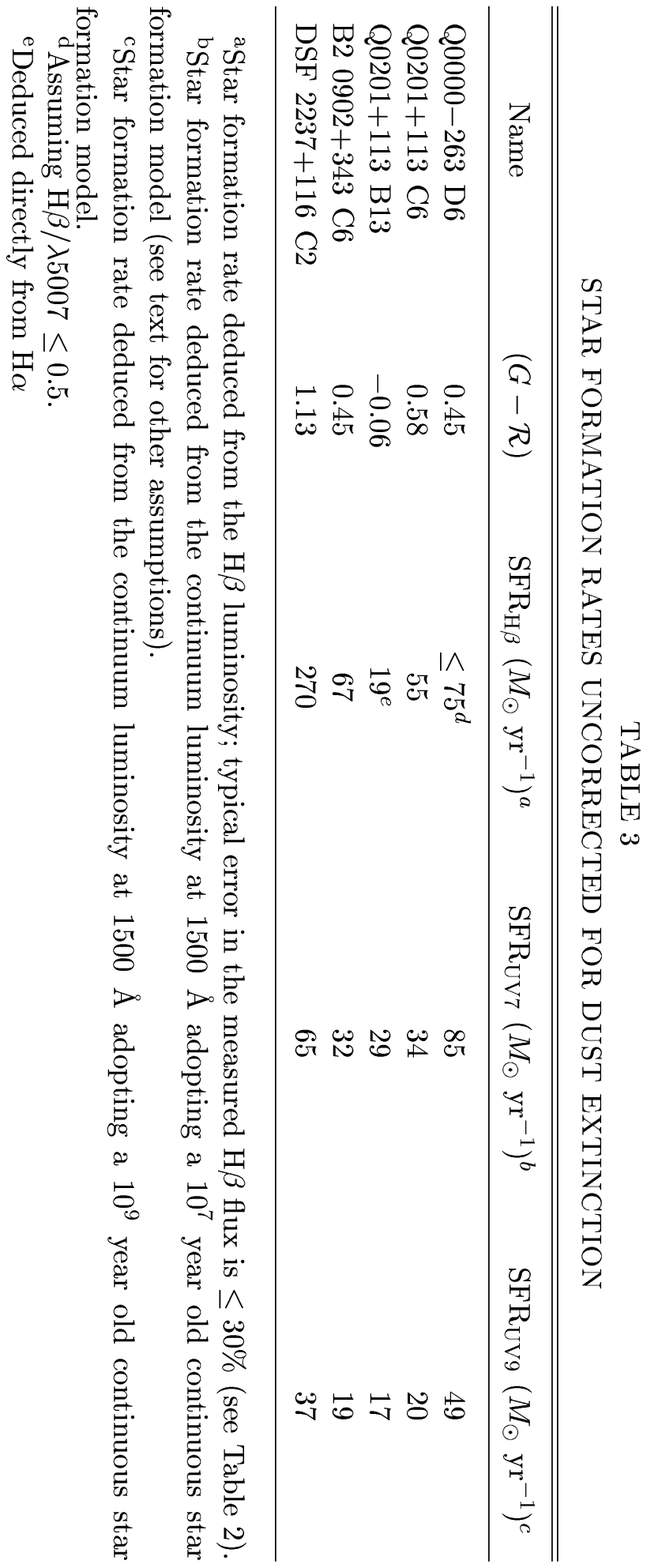}
\end{figure}


\hspace*{-5cm}
\begin{figure}
\vspace*{-1.5cm}
\figurenum{0}
\epsscale{1.1}
\plotone{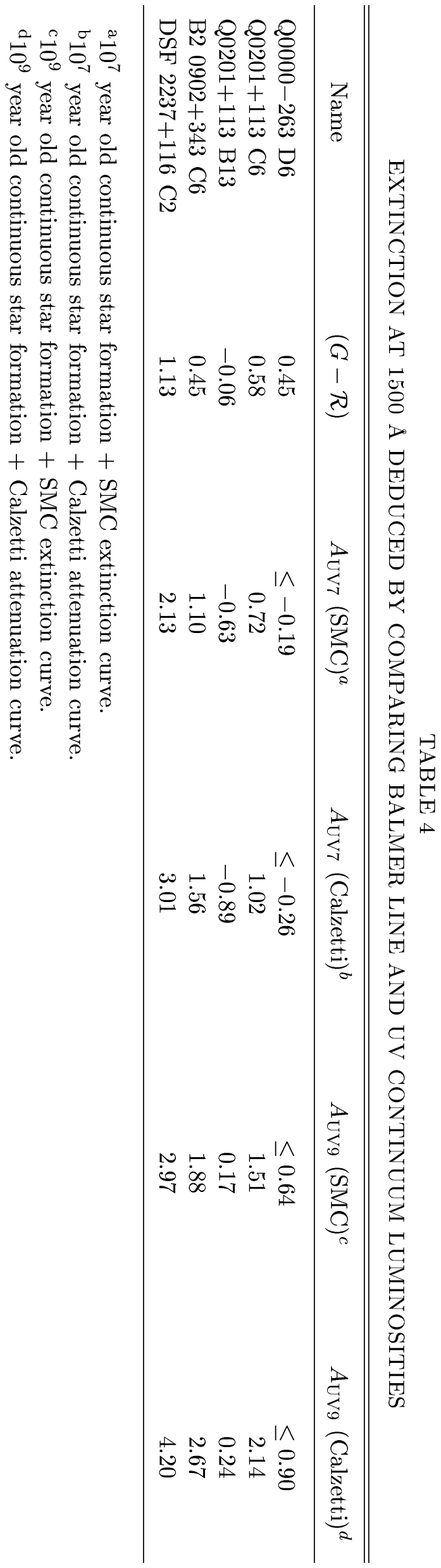}
\end{figure}

\hspace*{-5cm}
\begin{figure}
\vspace*{-1.5cm}
\figurenum{0}
\epsscale{1.2}
\plotone{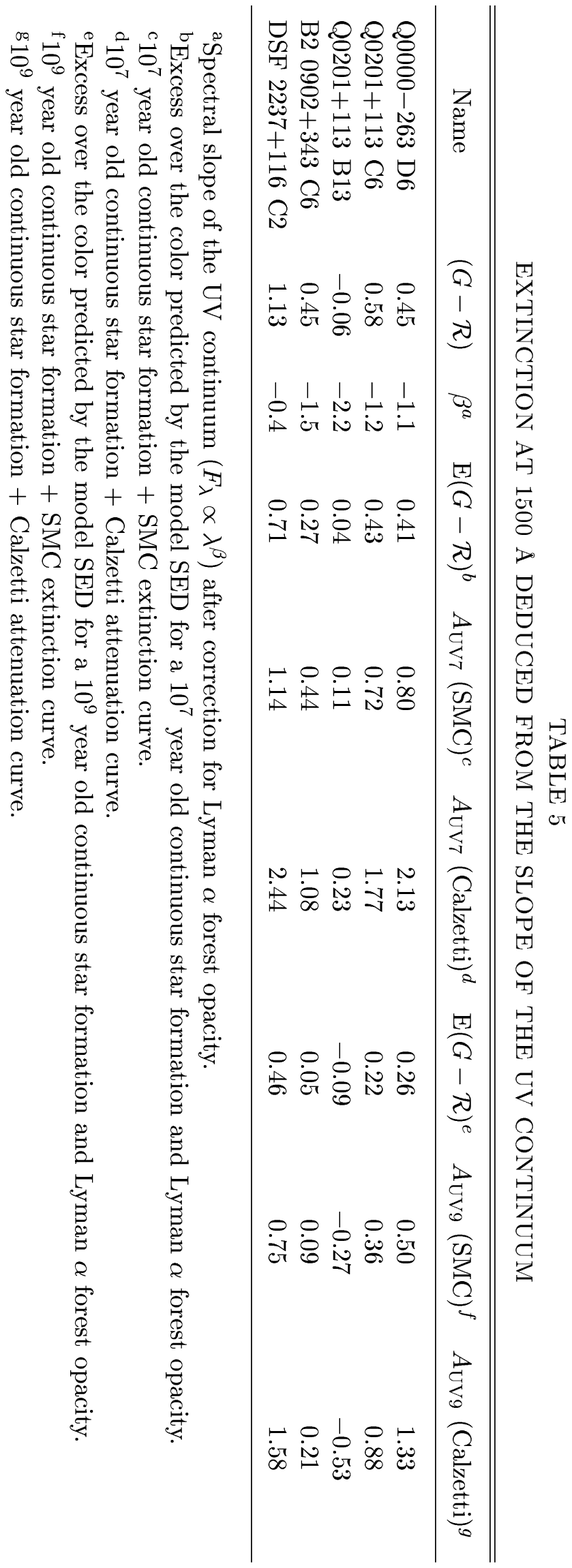}
\end{figure}

\addtocounter{table}{+5}

\begin{deluxetable}{lcccc}
\tablewidth{0pc}
\scriptsize
\tablecaption{VELOCITY DISPERSIONS AND VIRIAL MASSES}
\tablehead{
\colhead{Name} & \colhead{$\sigma$ (km~s$^{-1}$)} & 
\colhead{Half-light radius (arcsec)}& 
\colhead{Half-light radius (kpc)$^a$}&
\colhead{$M_{\rm vir}$ ($10^{10} M_{\odot}$)}  
}
\startdata
Q0000$-$263 D6     & $60 \pm 10$     & $0.22^b$       & 1.8     &  0.8      \nl
Q0201$+$113 C6     & $70 \pm 20$     & $0.25^b$       & 2.0     &  1.2      \nl
Q0201$+$113 B13    & $85 \pm 15$     & $0.2^c$        & 1.5     &  1.3      \nl
B2 0902$+$343 C6   & $55 \pm 15$     & \ldots         & \ldots  &  \ldots   \nl
DSF 2237$+$116 C2  & $190 \pm 25$   & $0.2^c$        & 1.5     &  5.5      \nl
\enddata
\tablenotetext{a}{$H_0 = 70$~km~s$^{-1}$; $q_0 = 0.1$}
\tablenotetext{b}{From {\it HST} NICMOS images (Giavalisco et al., in 
preparation)}
\tablenotetext{c}{From Keck II NIRC images}
\end{deluxetable}

\begin{deluxetable}{lccc}
\tablewidth{0pc}
\scriptsize
\tablecaption{RELATIVE VELOCITIES INTERSTELLAR 
ABSORPTION, NEBULAR EMISSION, AND LYMAN $\alpha$ EMISSION LINES}
\tablehead{
\colhead{Name} & \colhead{$V_{\rm IS~abs}$ (km~s$^{-1}$)} & 
\colhead{$V_{\rm H~II}$ (km~s$^{-1}$)} & 
\colhead{$V_{\rm Ly\alpha}$ (km~s$^{-1}$)}
}
\startdata
Q0000$-$263 D6     & $-435 \pm 50^a$    & 0         & $+375$            \nl
Q0201$+$113 C6     & $-3020 \pm 30$     & 0         & \ldots            \nl
Q0201$+$113 B13    & $+250$             & 0         & \ldots            \nl
B2 0902$+$343 C6   & $-800 \pm 300$   & 0         & $+575$            \nl
DSF 2237$+$116 C2  & $+110 \pm 60$    & 0         & $+1090$           \nl
\enddata
\tablenotetext{a}{The errors quoted are the standard deviation of 
different interstellar absorption lines.}
\end{deluxetable}

%
%

\newpage 
\begin{figure}
\figurenum{1}
\epsscale{1.1}
\hspace{-1.7cm}
\plotone{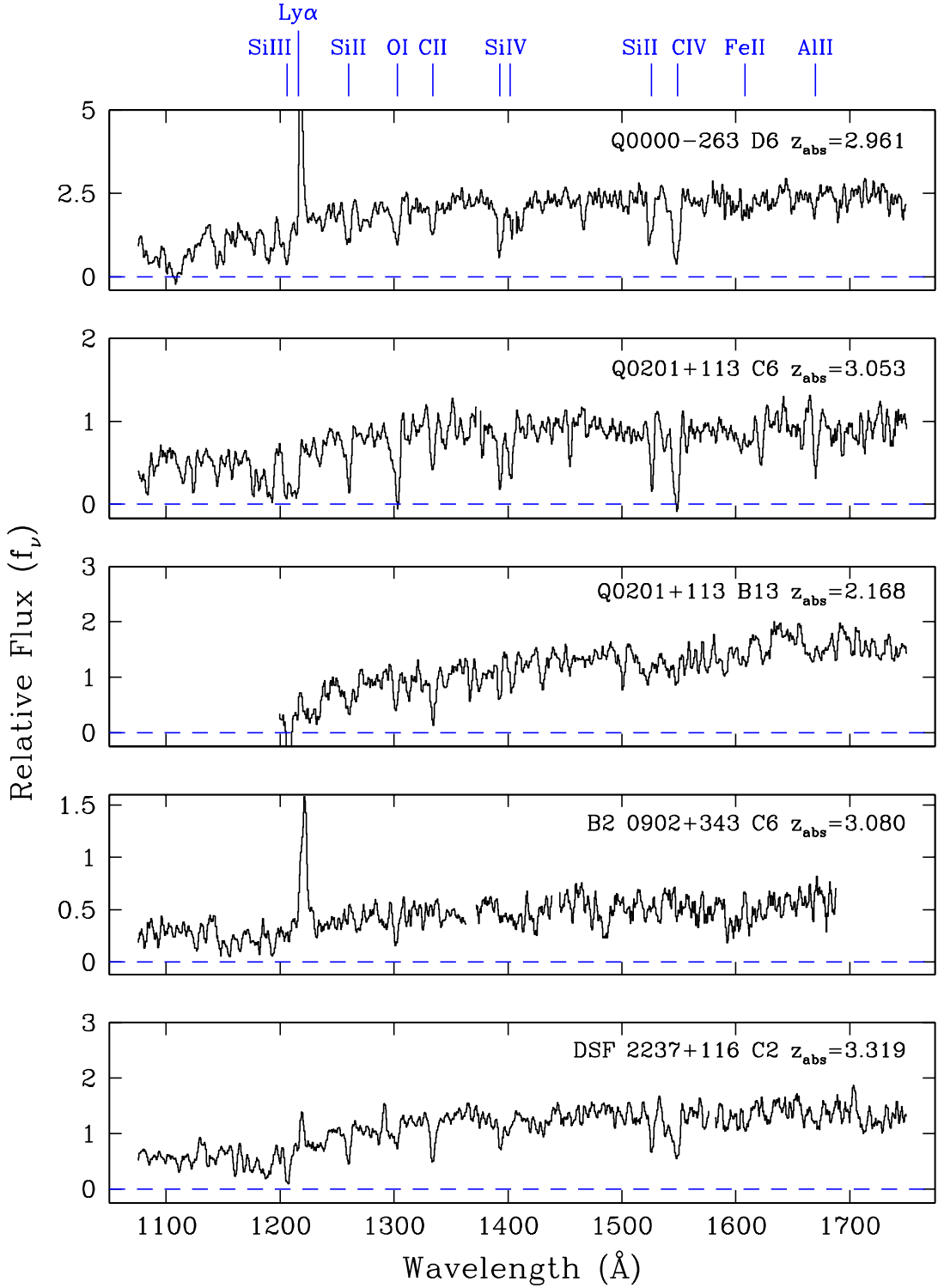}
\vspace{-3cm}
\caption{LRIS optical spectra of the five galaxies studied in this work.
Each spectrum is shown in the rest frame at the redshift of the interstellar
absorption lines; the flux scale is arbitrary. 
The positions of the 
most prominent interstellar lines are indicated at the top.
The spectra have been smoothed with a kernel of width 12 \AA\ (the spectral
resolution) for display.
} 
\end{figure}

\newpage 
\begin{figure}
\figurenum{2}
\epsscale{1.1}
\plotone{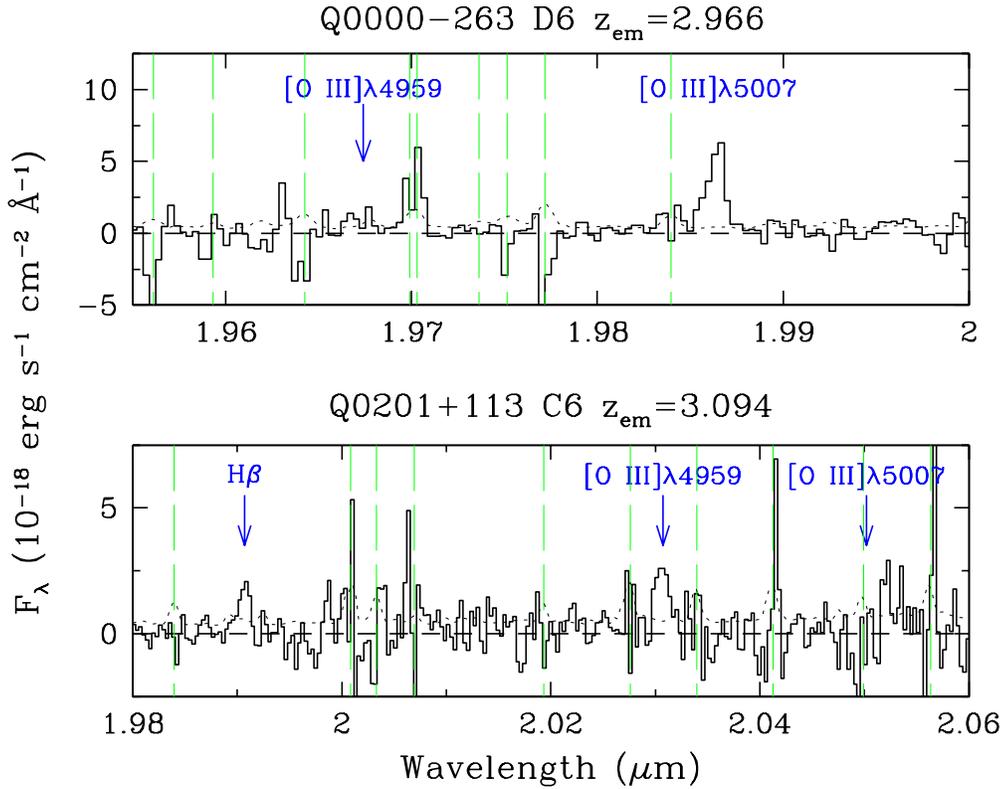}
\vspace{-7cm}
\caption{Portions of the {\it UKIRT} spectra of 
$z \simeq 3$ Lyman break galaxies
secured during the September 1996 observing run. 
The positions of the nebular emission lines covered are indicated. 
The short-dash line shows the $1 \sigma$ error applicable to each 
spectrum. The vertical long-dash lines mark the locations of the strongest
sky OH$^-$ emission features; although they have been subtracted out, 
large residuals can remain if the sky lines are saturated
(see text). 
} 
\end{figure}

\newpage
\begin{figure}
\hspace{-1.5cm}
\vspace{-3cm}
\figurenum{3}
\epsscale{1.1}
\plotone{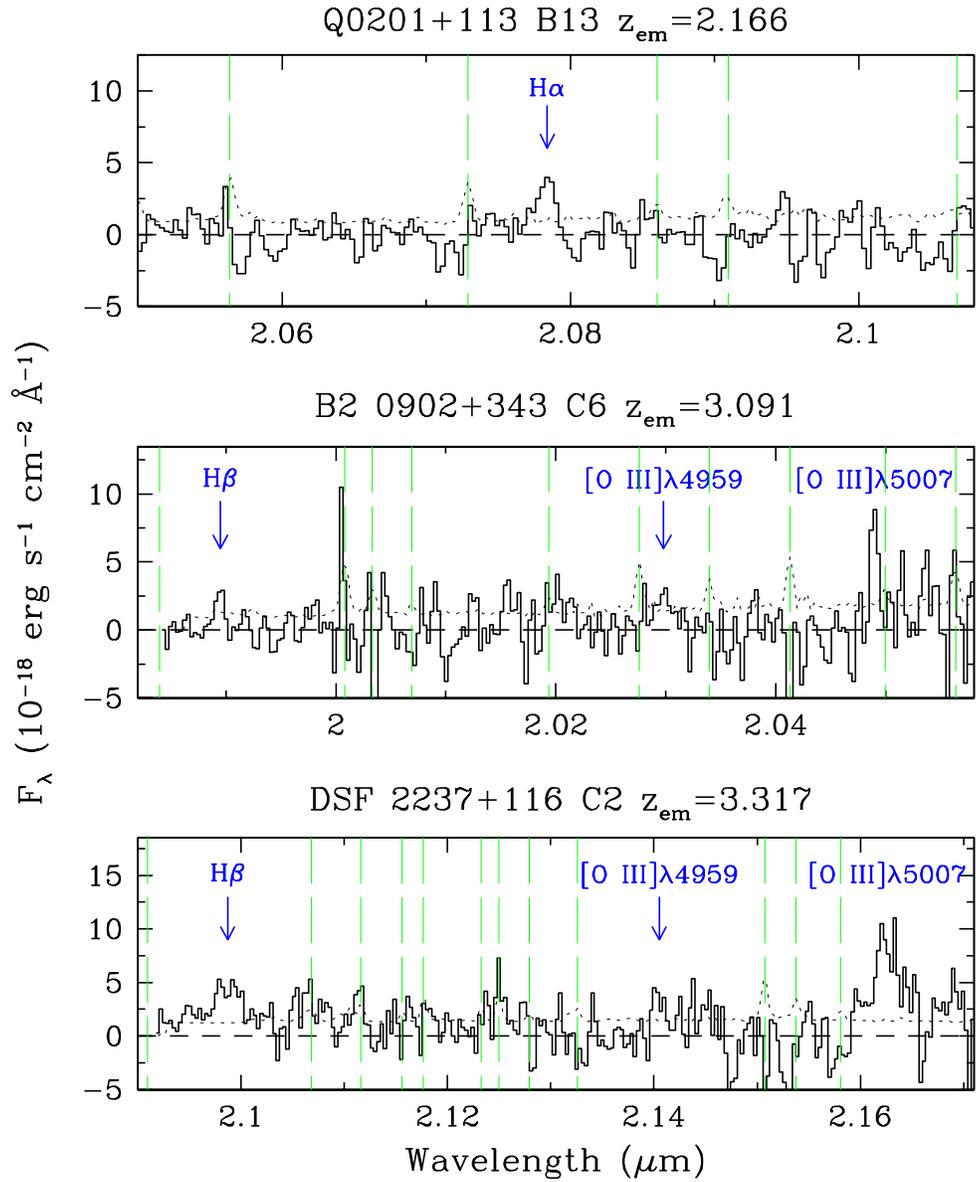}
\vspace{-3cm}
\caption{As for Figure 2, showing the spectra recorded in the 
1997 observing runs.
} 
\end{figure}

\newpage 
\begin{figure} 
\hspace{-1cm} 
\figurenum{4}
\epsscale{0.7} 
\plotone{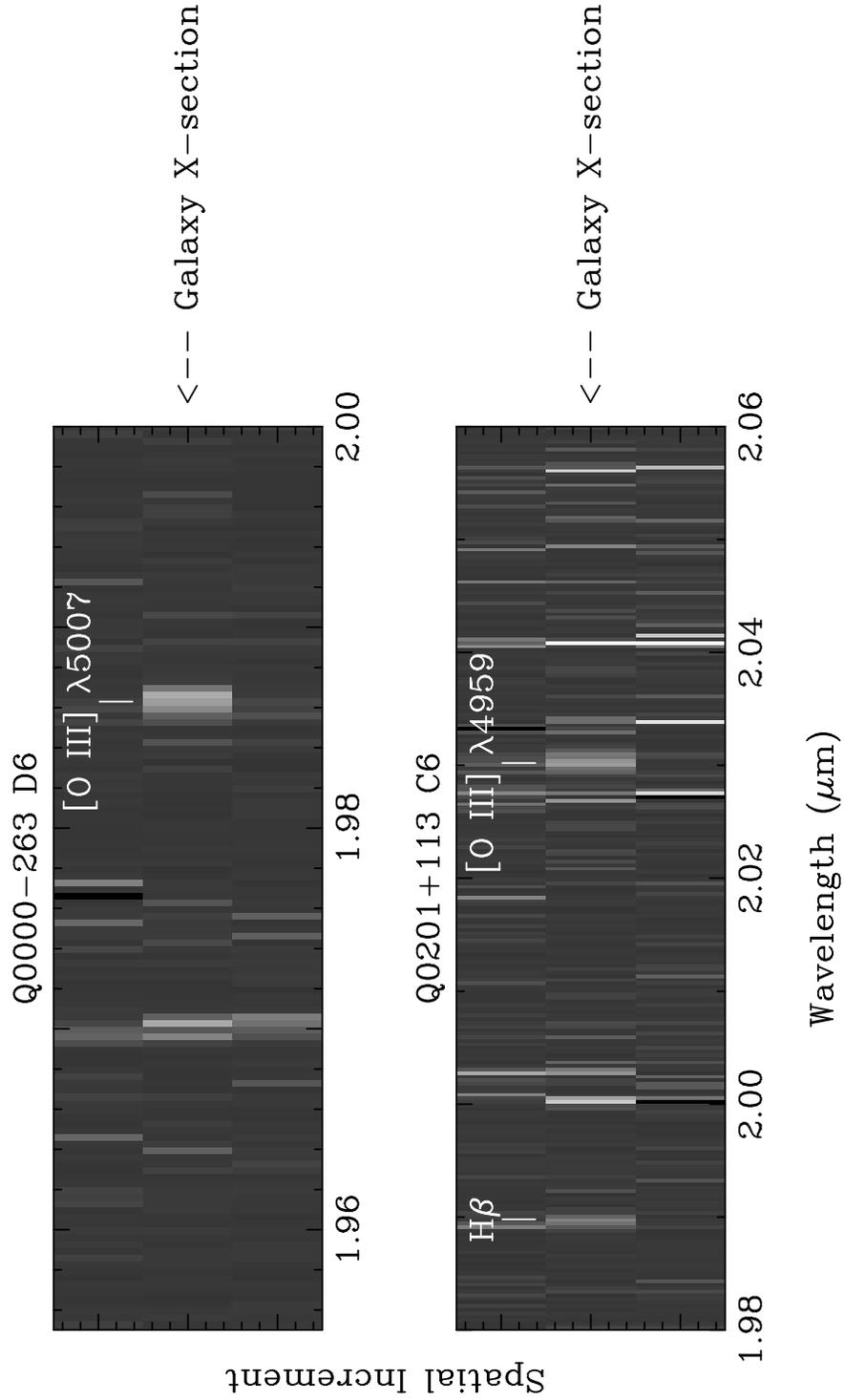} 
\vspace{1.2cm} 
\caption{Portions of
the co-added two dimensional CGS4 images of Q0000$-$263 D6 and Q0201$+$113 C6.
Each spatial increment is 1.2 arcsec along the slit; the position of each
galaxy on the slit was adjusted so that most of  the light is in the central
cross-section. Most of the bright pixels are residuals from the subtraction of
the strongest OH$^-$ sky emission lines, the positions of which are indicated 
in Figure 2.} 
\end{figure}

\newpage
\begin{figure}
\hspace{-1.8cm}
\figurenum{5}
\epsscale{1.1}
\plotone{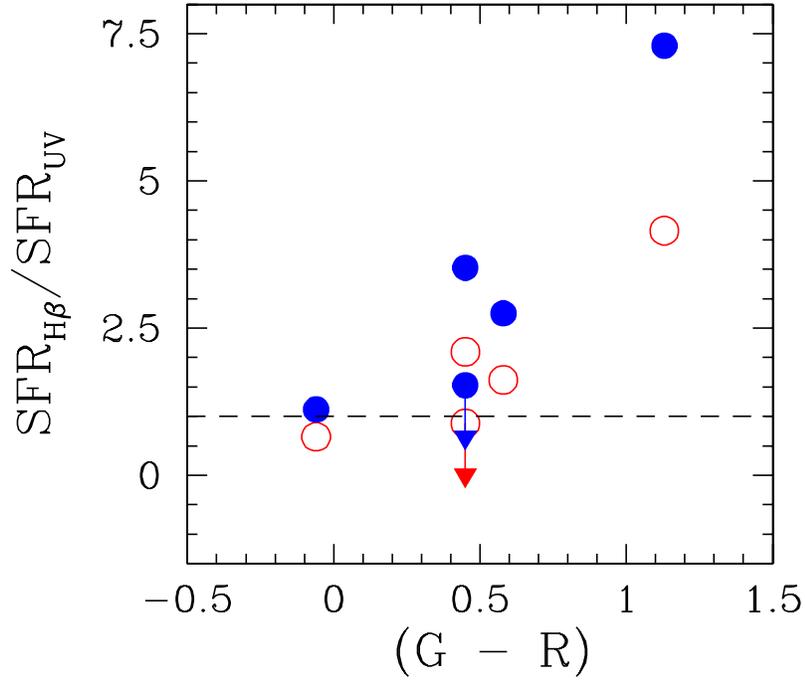}
\vspace{-6cm}
\caption{The ratio of the star formation rates implied by the luminosities
of the \hbeta\ line and of the UV continuum respectively is plotted as a function 
of the observed ($G - {\cal R}$) colour. Open circles: $10^7$ year old 
continuous star formation model; filled circles: $10^9$ year old 
continuous star formation model. The horizontal dashed line is at 
SFR$_{{\rm H}\beta}$ = SFR$_{\rm UV}$.
} 
\end{figure}

\end{document}